# Sub-Micrometer Particles Remote Detection in Enceladus' Plume Based on Cassini's UV Spectrograph Data

Jan Kotlarz, Katarzyna Kubiak
Research Network Łukasiewicz – Institute of Aviation, Al. Krakowska 110/114, 02-256 Warsaw, Poland

Natalia Zalewska
Space Research Center, Polish Academy of Sciences, Bartycka 18A, 00-716 Warsaw, Poland

**Abstract:** Enceladus is the Saturnian satellite is known to have water vapor erupting from its south pole region called „Tiger Stripes". Data collected by Cassini Ultraviolet Imaging Spectrograph during Enceladus transiting Saturn allow us to estimate water plume absorption from 1115.35–1912.50 Å and compare it to the Mie solutions of Maxwell equations for particles with a diameter in the range from 10 nm up to 2 μm. The best fit performed using Gradient Descent method indicates a presence of sub-micrometer particles of diameters: 120–180 nm and 240–320 nm consistent with Thermofilum sp., Thermoproteus sp., and Pyrobaculum sp. cell sizes present in hydrothermal vents on Earth.

## 1. Introduction

Enceladus is a very interesting Saturn's mid-size moon in the field of astrobiology. After the discovery of water vapour plumes erupting from its south pole region [16] made by Cassini probe during a flyby in July 2005, data collected by all instruments were investigated. Particularly useful for plumes composition estimation were studies based on solar and stellar occultations [6, 12, 4, 13], but also transit passing the Saturn's ring [5]. Ice grain size distribution has been determined using Cassini's Dust Analyzer (CDA) data. The least ice particles ($< 0.4$ $\mu$m) were observed reaching escape velocity and going into space and in the result appearing Saturn's E-ring. Those particles were interpreted as "condense from gas in the plume" [4], unlike larger ice grains that were interpreted as formed from matter originated from the bottom of the ocean [14]. Charged nanograins in a range of diameter between 2.2 nm and 3.4 nm were reported by Hill et al. [8]. Also sub-micron ice particles presence on Enceladus surface, possibly derived as plume deposits was described by Scipioni et al. [15].

Particles smaller than 0.5 $\mu$m correspond to the size of single cells of thermophilic bacteria and archaea living inside Earth's hydrothermal vents with temperatures near 80 °C. Thermophilic cells are smaller than typical 1–2 $\mu$m microorganisms. The smallest cell sizes recognized in hyperthermophilic archaea are 0.17 $\mu$m in diameter (*Thermofilum* sp.), 0.3 $\mu$m in diameter (*Thermoproteus* sp. and *Pyrobaculum*), or disks 0.2–0.3 $\mu$m in diameter and 0.08–0.1 $\mu$m wide in *Thermodiscus* and *Pyrodictium*. The presence of methanogens in the ocean of Enceladus [10, 18] would result in the presence of particles in water plumes of sub-micrometer size, consistent with the diameter of the cells.

Estimating particle size distribution on spectral data analysis is a common practice. One method is to compare the observed brightness of the scattering molecules with the Mie solutions of light scattering for Maxwell's equations. For example Hedman et al. in 2009 have estimated particle size distribution for $> 1$ μm diameters using a.o. Mie results for 160° scattering angle [7]. Gao et al. in 2016 have estimated monomer and mean particle diameter distribution using the same method [3].

The main goal of our work was to confirm the presence of the sub-micrometer particles using far-ultraviolet part of the Cassini Ultraviolet Imaging Spectrograph Subsystem (UVIS) spectrum acquired during Enceladus passing in front of their parent planet in 2009 and compare results with known methanogenic archaea and bacteria sizes investigated by taking samples from hot (up to 80 °C) geysers on the Earth.

## 2. Materials and methods

To describe the sub-micrometer particle size distribution in Enceladus plumes, we used ultraviolet hyperspectral data collected by the Cassini probe in 2009. Using this data, we calculated the ratio of the Saturn's signal disturbed by the Enceladus transit to the Saturn's pure signal. Then, using Mie solutions of Maxwell's equations, we estimated effective cross-sections for particles with diameters between 10 nm and 2 $\mu$m. Effective cross-sections were estimated for 1024 UV wavelengths measured by Cassini UVIS. Using the gradient-descent method, we estimated particle diameter distribution in plumes assuming that the modeled ratio of the disturbed to undisturbed signal should coincide with the spectrum observed by UVIS.

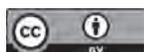

## 2.1. Cassini's UVIS data and scattering model

To estimate pure Saturn's disc UV signal and the impact of Enceladus plumes, we have used UVIS observation of Enceladus transit of Saturn on 2 November 2009. Far ultraviolet measurements were made by positioning spectrometer's slit on Saturn's disc, just below Enceladus transit trajectory. The far-ultraviolet (FUV) high resolution slit with 1024 spectral bands 1115.35–1912.50 Å records unocculted and occulted spectrum of Saturn including Ly-$\alpha$ line. Based on this observations for each band we estimated ratio between unocculted $j_{un}$ and occulted signal $j_{oc}$. Ratio in the form $1 - j_{oc}/j_{un}$ convenient for further computations is shown on the Figure 1 (black solid line). We assumed that in the volume unit of Enceladus' plume there are $N$ particles (discs) that can scatter the radiation reflected from Saturn. $j_{un}$ denote the incident photon flux (the number of photons passing per unit time through a unit area of plume), and X the number of scattering processes per unit volume per unit time. Cross section $\sigma$ is defined by the equation 1.30 in Kubiak [9]:

$$j_{un}N = X \qquad (1)$$

$j_{un}$ value is the unocculted Saturn's signal and was measured on 2 November 2009 by Cassini's UVIS spectrometer before Enceladus transit [5]. Let $d$ denote plume (scattering matter) thickness. Then $Nd$ would be the number of scattering particles between Saturn and Cassini in the unit area of 1 m². At the other hand $j_{oc} = j_{un} - Xd$ would be occulted Saturn's signal and:

$$j_{un}Nd\sigma = Xd = j_{un} - j_{oc} \qquad (2)$$

Finally, the equation with a ratio between occulted and unocculted signal is expressed mathematically by a factor of particles cross section and the number of scattering particles between Saturn and Cassini in the unit area:

$$\sigma Nd = 1 - \frac{j_{oc}}{j_{un}} \qquad (3)$$

Right-hand side of the equation 3 can be estimated using UVIS data, left-hand side is the function of the scattering cross-sections of particles and the number of scattering particles between Saturn and Cassini in the unit area. In the approach presented in this study we assumed that for a mix of particles with different diameters left-hand side of the equation 3 is a sum of cross-sections multiplied by the number of particles for each component. The equation 3 is calculated for each wavelength apart.

## 2.2. Mie solutions of Maxwell equations for 180º

We calculated Mie solutions of Maxwell equations using the Python Mie Scattering package (PyMieScatt) [17]. Using Cassini's Imaging Science Subsystem (ISS) data we estimated the angular diameter of the plumes taking into account the spacecraft's distance from Enceladus on the day of the moon's transit. The assumption is made that the recorded signal comprises both the non-scattered signal from Saturn and the radiation scattered at a 180° angle. A spectrometer is an instrument used to measure the intensity of light at different wavelengths in a spectrum, providing valuable information

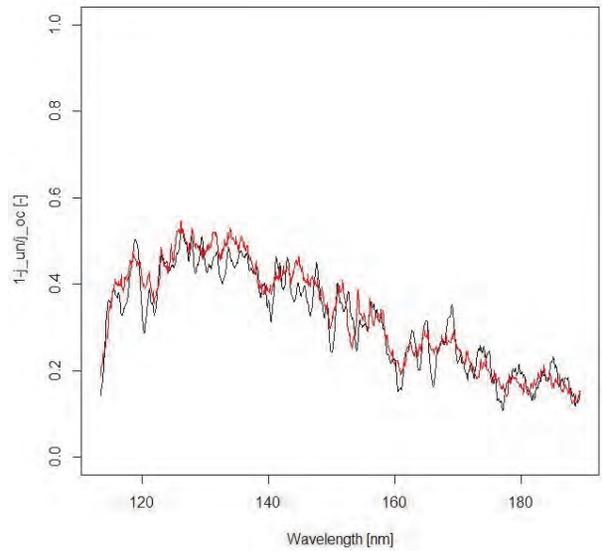

**Fig. 1. Black solid line** – ratio of Saturn's ultraviolet signal $j_{un}$ to signal $j_{oc}$ disturbed by Enceladus plumes computed using Cassini's UVIS data (1 – $j_{oc}/j_{un}$, RHS of the Eq. 3). Red solid line – scattering of the result mixture (see Figure 3) of particles with diameters between 10 nm and 2 $\mu$m at an angle of 180°
Rys. 1. Czarna linia ciągła – stosunek sygnału ultrafioletowego Saturna $j_{un}$ do sygnału $j_{oc}$ zakłóconego przez pióropusze Enceladusa obliczony przy użyciu danych UVIS sondy Cassini (1 – $j_{oc}/j_{un}$, prawa strona równania 3). Czerwona linia ciągła – rozproszenie otrzymanej mieszaniny (zob. Rys. 3) cząstek o średnicach od 10 nm do 2 $\mu$m pod kątem 180° about the composition and properties of the observed sources. What adds complexity is that the cross-section of the plume particles is not solely determined by their diameter but also by the relationship between the wavelength of the scattered light and the diameter of these particles. This intricacy underscores the significance of the size-wavelength ratio in the applicability of Mie theory. The theory's utility hinges on its ability to account for such nuanced interactions between particle size and the wavelength of scattered light, making it a valuable tool in understanding complex phenomena like those observed in the study. To model this effect, based on Mie solutions we computed cross sections of a single molecule over the wavelength range of 1115.35–1912.50 Å every 0.79 Å (1024 values consistent with UVIS bands). The cross section values were computed for particles with diameters of 10–1000 nm (every 10 nm), 1100–2000 nm (every 100 nm).

## 2.3. Gradient Descent technique

Left-hand side of the Equation 3 depends on the particles density in the plume, scattering layer depth and particles cross-sections. Right-hand side of this equation is given by Cassini's data. Such an approach enables the determination of the relative proportion of particles within a specified size range based on the observation of the scattered light intensity and solutions provided by Mie theory. This method allows for the application of established numerical techniques that fit theoretical models to the observed phenomenon. By utilizing known numerical methods for model fitting, researchers can extract valuable insights into the density of particles in the plume, the depth of the scattering layer, and the cross-sections of the particles. This integrative approach, combining observational data and theoretical frameworks, enhances the understanding of complex atmospheric phenomena and contributes to the refinement of our knowledge about the composition and behavior of plume particles. Our next step was to estimate $Nd$ values for particles with diameters between 10 nm and 2 $\mu$m. We used here method known as gradient descent technique [19]. For each type of par-ticles described by their diameter we defined it's number $N_i$ in the unit volume of plume as the unknown feature. There were 110 particle types: 100 types with sub-micrometer diameters (10 nm – 1 $\mu$m, every 10 nm) and 10 types with diameters over 1 $\mu$m (1100–2000 nm, every 100 nm). Than we hypothesize:

$$h_N = N_0 + \sum_{i=1}^{110} N_i \sigma_i(\lambda). \qquad (4)$$

where $N_0$ is a bias, $\sigma_i$ is a cross section computed with Mie solution for particle with $i$-th diameter and wavelength $\lambda$. Let's note, that $h_N d$ is equal to left-hand side of the equation 3, so $1 - j_{oc}/j_{un}$ may be used as a known value. In the beginning of each approach we were setting random $N_i$ values (initial mixed state). Then for $N_i$ values $h_N$ were calculated. In the next step we were modifying iterative $N_i$ values to minimize cost function:

$$J = \left( h_N - \left(1 - \frac{j_{oc}}{j_{un}}\right) \right)^2 \qquad (5)$$

until $\frac{dJ}{dN} < \varepsilon$ where $\varepsilon$ is a small value ($\approx 10^{-4}$). After this algorithm returned best-fit mixture of scattering particles. We have repeated this procedure $10^4$ times with different mixed states. Our best-fit mixed stated (final mixed state) was returned as a description of the compounds of the plume.

## 3. Results

Ratio of the occulted and unocculted Saturn's FUV signal in the form of $1 - j_{oc}/j_{un}$ is presented on the Figure 1 (black solid line). Example Cost function value vs the number of iteration is presented in the Figure 2. Best-fit scattering particles layer (plume) composition is presented in the Figure 3 by black points. Result values processed with the Savitzky-Golay filter is presented by red solid line. Best-fit mixed state's ratio between modeled scattering signature and Saturn's signal is presented in the Figure 1 by red line.

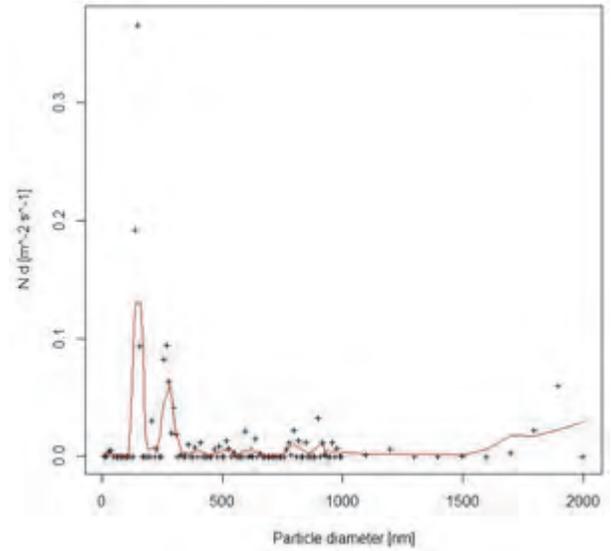

**Fig. 3. Best estimation of the scattering mixture components. Black points – exact values representing the amount of scattering particles from 10 nm to 2 $\mu$m in diameter. Red line – the same data processed by the Savitzky-Golay filter. The result indicates the presence of particles with a diameter of 120–180 nm, 240–320 nm**
Rys. 3. Najlepsze oszacowanie składników mieszaniny rozpraszającej. Czarne punkty – dokładne wartości reprezentujące ilość rozpraszających cząsteczek o średnicy od 10 nm do 2 $\mu$m. Czerwona linia – te same dane przetwarzane przez filtr Savitzky'ego-Golaya. Wynik wskazuje na obecność cząsteczek o średnicy 120–180 nm, 240–320 nm

## 4. Discussion

The ratio of ultraviolet wavelengths to the diameters of sub-micrometer particles is suitable for scattering mater component detection using Mie solutions of Maxwell equations. The dimensionless size parameter $x = 2\varpi r/\lambda$ connecting the particle radius $r$ and the wavelength of the incident electromagnetic radiation is consistent with the ranges used for this parameter in the literature [17]. Comparison between Cassini's FUV range (1115.35–1912.50 Å) gives us $x$ values between 0.3 and 57.1 for sub-micrometer particles. The most widely applied numerical Mie code from Bohren and Huffman [2] is restricted to size parameters $< 2 \cdot 10^4$ [21] which makes this approach suitable for detecting small molecules both with instruments in the UV range and even with visible light. Gradient descent technique used in our approach to estimate particle size distribution in the plume was selected because of the computational complexity $O(kn^2)$ where $k$ is the number of iterations and n is the number of features.

Mie theory is suitable for a problem discussed in this article. Wang et al. [20] discusses the investigation of Poynting vector field lines around small particles based on classical Mie theory. It highlights that particles can efficiently absorb energy near optical resonance, where their optical absorption cross-section surpasses the geometrical cross-section. The concept of an "input window" is introduced, indicating that absorbed energy flows through a limited portion of the particle's surface. This window expands with the increasing imaginary part of the particle's dielectric function. For small values, absorbed energy is released through plasmon radiation, creating complex energy flux patterns in the near-field region. Authors emphasizes the inadequacy of the dipole approximation and suggests considering higher-order terms in size parameter $q \sim 2/l$ where $l$ is the wavelength. The applicability of Mie solutions is broad, particularly in studying microscopic energy absorption phenomena, with potential applications in nanotechnology, biophysics, and material sciences. However, practical implementation for $q/l$

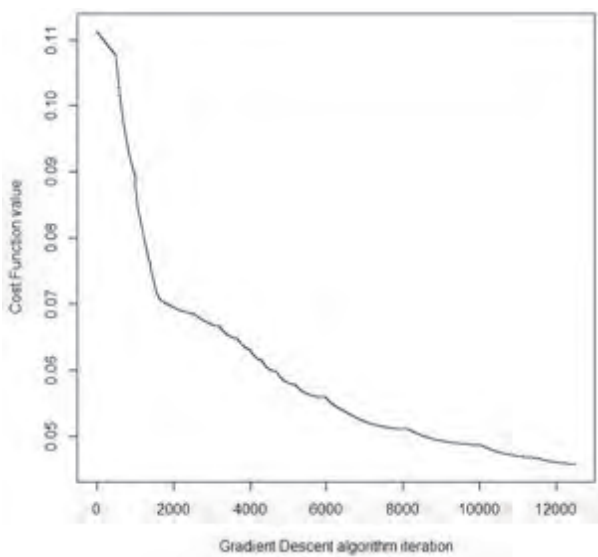

**Fig. 2. Example Cost Function of mixture evolution during Gradient Descent algorithm. Cost Function represents the difference between data measured by Cassini's UVIS and modeled by Mie Solution of Maxwell equations for the mixture**
Rys. 2. Przykładowa ewolucja funkcji kosztu mieszaniny podczas algorytmu Gradient Descent. Funkcja kosztu reprezentuje różnicę między danymi zmierzonymi przez UVIS sondy Cassini a modelowanymi przez rozwiązania Mie równań Maxwella dla mieszaniny wynikowej values extending boundary values $0.1 < q/l < 10.0$ may require considering more advanced effects beyond Mie theory's scope.

Because of the random mix $\vec{N}$ generated at the beginning of the procedure $h_N(\vec{N})$ might be different in order of magnitude that function $1 - j_{oc}/j_{un}$. For this reason in first iterations $dJ(\vec{N})/dN_i$ were significantly smaller or larger than zero for each feature $i$. After general scaling $h_N$ to this function the process of eliminating mismatched components began. These phases can be seen on the cost function $J$ vs iteration plot (Fig. 2). During first 1,800 iteration we can see general $h_N$ scaling resulting fast cost function reduction from 0.11 to 0.07. Further cost function reduction from 0.07 to values below 0.05 has been made during 11,000 iterations. We can see characteristic saw-tooth component of the function. Each iteration when cost function starts to diminish suddenly is characterized by the elimination of particular component $j$ and setting $N_j = 0$ for next iterations.

Computational complexity depends on the number of components (features) like $n^2$, so the final iterations were carried out quickly ($110^2/n^2_{final}$ faster than first iteration). This property of the gradient descent technique allowed achieve a cost function minimum with an accuracy of up to $10^{-4}$ in the most of the random mixed states. The result best-fit estimation of sub-micrometer particle diameters distribution (Fig. 3, red line) give us two types of particles: characterized by diameters of 120–180 nm and 240–320 nm. Besides these two types we can see particles with diameters up to $1\,\mu m$ with a population of about 10–20 times lower than two main components. Also the micrometer-size particles discovered in many studies [11] are present in our result. The need expressed by Bedrossian et al. in 2017 [1] that "detection of extant microbial life (…) requires the ability to identify and enumerate micrometer-scale, essentially featureless cells" could be satisfied by ultraviolet measurement done by occultations.

# Detekcja cząstek submikrometrowych w pióropuszu Enceladusa na podstawie danych ze spektrografu UV sondy Cassini

**Streszczenie:** Enceladus, księżyc Saturna, jest charakterystyczny ze względu na erupcje pary wodnej z regionu jego bieguna południowego, tzw. „Tiger Stripes". Dane zebrane przez instrument sondy Cassini: Ultraviolet Imaging Spectrograph podczas tranzytu Enceladusa przed tarczą Saturna pozwalają oszacować absorpcję światła przez pióropusze wodne w zakresie 1115,35–1912,50 Å i porównać ją z rozwiązaniami Mie równań Maxwella dla cząsteczek o średnicach w zakresie od 10 nm do 2 μm. Najlepsze dopasowanie wykonane metodą Gradient Descent wskazuje na obecność cząstek sub-mikrometrowych o średnicach: 120–180 nm i 240–320 nm zgodnych z rozmiarami komórek Thermofilum sp., Thermoproteus sp. i Pyrobaculum sp. obecnych w kominach hydrotermalnych na Ziemi.

**Słowa kluczowe:** astrobiologia, misje kosmiczne, teledetekcja, rozpraszanie Mie, Enceladus

### Katarzyna Kubiak, PhD
Katarzyna.Kubiak@ilot.lukasiewicz.gov.pl
ORCID: 0000-0002-4156-3139

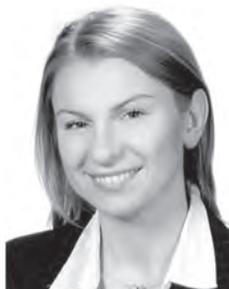

She has graduated from the Warsaw University of Life Sciences. Since 2015 she has been working as an assistant professor at the Department of Remote Sensing, Łukasiewicz – Institute of Aviation and currently works with the acquisition and analysis of spectral data (laboratory and field spectrometry, UAV, satellites). The main research interests are the optimisation of the configuration of optical sensors of multispectral cameras for imaging biomass and natural components.

### Jan Kotlarz, MSc
Jan.Kotlarz@ilot.lukasiewicz.gov.pl
ORCID: 0000-0002-8212-7798

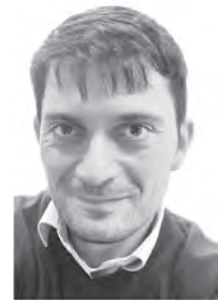

Graduate in astronomy from the Faculty of Physics at the University of Warsaw, in Management and in the Environmental Protection from the WSB Merito University in Poznań. Since 2012, employed at the Institute of Aviation. Currently pursuing a PhD at the Nicolaus Copernicus University in Toruń.

### Natalia Zalewska, PhD
natalia@cbk.waw.pl
ORCID: 0000-0001-8843-4396

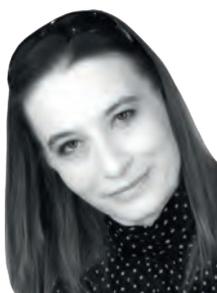

She works at the Space Research Center of the Polish Academy of Sciences; she is a specialist in the field of Mars geological research conducted based on data from the Mars Express and MRO satellites. In 2005 and 2019, she took part in similar missions on this planet, organized by the Mars Society, which took place at the Martian base in the Utah desert. Currently, he is involved in the search for water and conical volcanoes on the surface of Mars. She is involved in the popularization of space through cooperation with the media.